\documentclass[11pt,preprint2]{aastex}

\def\pc{\ifmmode \mathrm{pc} \else $\mathrm{pc}$ \fi}
\def\mpc{\ifmmode \mathrm{Mpc} \else $\mathrm{Mpc}$\fi}
\def\mpcthree{\ifmmode \mathrm{Mpc}^{-3} \else $\mathrm{Mpc}^{-3}$\fi}
\def\gpcthree{\ifmmode \mathrm{Gpc}^{-3} \else $\mathrm{Gpc}^{-3}$\fi}

\def\kelvin{\ifmmode \mathrm{K} \else {$\mathrm{K}$}\fi}
\def\kev{\ifmmode \mathrm{keV} \else $\mathrm{keV}$ \fi}

\def\lsun{\ifmmode {L_\odot} \else $L_\odot$\fi}
\def\msun{\ifmmode M_\odot \else $M_\odot$\fi}
\def\msunyr{\ifmmode M_\odot~\mathrm{yr}^{-1} \else $M_\odot~\mathrm{yr}^{-1}$\fi}

\def\mgii{\ifmmode {\rm Mg{\sc ii}} \else Mg~{\sc ii}\fi}
\def\civ{\ifmmode {\rm C{\sc iv}} \else C~{\sc iv}\fi}
\def\civ{\ifmmode {\rm C{\sc iv}} \else C~{\sc iv}\fi}
\def\mgii{\ifmmode {\rm Mg{\sc ii}} \else Mg~{\sc ii}\fi}
\newcommand{\oiii}{{\sc [O~iii]}}

\newcommand{\nev}{{[Ne~{\sc v}]}}

\newcommand{\feii}{Fe~{\sc ii}}

\def\teff{\ifmmode {T_{\rm eff}} \else $T_{\rm eff}$\fi}
\def\tmax{\ifmmode {T_{\rm max}} \else $T_{\rm max}$\fi}

\def\mbh{\ifmmode {M_{\rm BH}} \else $M_{\rm BH}$\fi}
\def\led{\ifmmode L_{\mathrm{Ed}} \else $L_{\mathrm{Ed}}$\fi}
\def\lbol{\ifmmode L_{\mathrm{bol}} \else $L_{\mathrm{bol}}$\fi}
\def\mdot{\ifmmode {\dot M} \else $\dot M$\fi}
\def\mdoto{\ifmmode {\dot{M}_0} \else  $\dot{M}_0$\fi}

\def\hnot{\ifmmode H_0 \else H$_0$ \fi}

\def\vkep{\ifmmode v_\mathrm{Kep} \else $v_\mathrm{Kep}$ \fi}
\def\vc{\ifmmode v_\mathrm{c} \else $v_\mathrm{c}$ \fi}

\def\vthree{\ifmmode v_{1000} \else $v_{1000}$ \fi}
\def\vkick{\ifmmode v_\mathrm{kick} \else $v_\mathrm{kick}$ \fi}
\def\eflare{\ifmmode E_\mathrm{flare} \else $E_\mathrm{flare}$ \fi}
\def\tflare{\ifmmode t_\mathrm{flare} \else $t_\mathrm{flare}$ \fi}
\def\lflare{\ifmmode L_\mathrm{flare} \else $L_\mathrm{flare}$ \fi}
\def\fflare{\ifmmode F_\mathrm{flare} \else $F_\mathrm{flare}$ \fi}
\def\nflare{\ifmmode N_\mathrm{flare} \else $N_\mathrm{flare}$ \fi}
\def\tshock{\ifmmode T_\mathrm{shock} \else $T_\mathrm{shock}$ \fi}
\def\rbound{\ifmmode R_\mathrm{b} \else $R_\mathrm{b}$ \fi}
\def\mbound{\ifmmode M_\mathrm{b} \else $M_\mathrm{b}$ \fi}
\def\tbound{\ifmmode t_\mathrm{b} \else $t_\mathrm{b}$ \fi}
\def\tagn{\ifmmode t_\mathrm{AGN} \else $t_\mathrm{AGN}$ \fi}
\def\rlim{\ifmmode R_\mathrm{lim} \else $R_\mathrm{lim}$ \fi}
\def\vlim{\ifmmode v_\mathrm{lim} \else $v_\mathrm{lim}$ \fi}
\def\mlim{\ifmmode M_\mathrm{lim} \else $M_\mathrm{lim}$ \fi}
\def\tlim{\ifmmode t_\mathrm{lim} \else $t_\mathrm{lim}$ \fi}
\def\llim{\ifmmode L_\mathrm{lim} \else $L_\mathrm{lim}$ \fi}
\def\vthree{\ifmmode v_{1000} \else $v_{1000}$ \fi}

\def\hbeta{\ifmmode \rm{H}\beta \else H$\beta$\fi}
\def\hbetan{\ifmmode \rm{H}\beta_{\rm n} \else H$\beta_{\rm n}$\fi}
\def\lalpha{\ifmmode \rm{Ly}\alpha \else Ly$\alpha$}

\def\dvhb{\ifmmode \Delta v_{\hbeta} \else $\Delta v_{\hbeta}$\fi}
\def\dvmg{\ifmmode \Delta v_{\rm{Mg}} \else $\Delta v_{\rm{Mg}}$\fi}

\def\yr{\ifmmode {\rm yr} \else  yr \fi}
\def\kms{km s$^{-1}$}
\def\cm{\ifmmode {\rm cm} \else  cm \fi}
\def\cmmitwo{\ifmmode \rm cm^{-2} \else $\rm cm^{-2}$\fi}
\def\cmmithree{\ifmmode \rm cm^{-3} \else $\rm cm^{-3}$\fi}
\def\cmps{\ifmmode \rm cm~s^{-1}\else $\rm cm~s^{-1}$\fi}
\def\cmpsps{\ifmmode \rm cm~s^{-2}\else $\rm cm~s^{-2}$\fi}
\def\kmps{\ifmmode \rm km~s^{-1}\else $\rm km~s^{-1}$\fi}
\def\kmpspmpc{\ifmmode \rm km~s^{-1}~Mpc^{-1} \else
    $\rm km~s^{-1}~Mpc^{-1}$\fi}
   
\def\erg{\ifmmode {\rm erg} \else $\rm erg$ \fi}
\def\ergps{\ifmmode {\rm erg~s^{-1}} \else $\rm erg~s^{-1}$ \fi}
\def\ergpspcm{\ifmmode \rm erg~s^{-1}~cm^{-2} \else $\rm erg~s^{-1}~cm^{-2}$ \fi}
\def\ergpspcmphz{\ifmmode \rm erg~s^{-1}~cm^{-2}~Hz^{-1} \else $\rm
erg~s^{-1}~cm^{-2}~Hz^{-1}$ \fi}
\def\ergpspcmpa{\ifmmode \rm erg~s^{-1}~cm^{-2}~\AA^{-1} \else $\rm
erg~s^{-1}~cm^{-2}~\AA^{-1}$ \fi}
\def\ergpsphz{\ifmmode \rm erg s^{-1} Hz^{-1} \else
   $\rm erg s^{-1} Hz^{-1}$ \fi}

\def\lam{\ifmmode {\lambda} \else {$\lambda$} \fi}
\def\llam{\ifmmode {L_\lambda} \else  $L_\lambda$ \fi}
\def\lamLlam{\ifmmode \lambda L_{\lambda}(5100) \else {$\lambda L_{\lambda}(5100)$} \fi}
\def\nuLnu{\ifmmode \nu L_{\nu}(5100) \else {$\nu L_{\nu}(5100)$} \fi}
\def\ilam{\ifmmode {I_\lambda} \else  $I_\lambda$ \fi}
\def\flam{\ifmmode {F_\lambda} \else  $F_\lambda$ \fi}
\def\inu{\ifmmode {I_\nu} \else  $I_\nu$ \fi}
\def\fnu{\ifmmode {F_\nu} \else  $F_\nu$ \fi}
\def\bnu{\ifmmode {B_\nu} \else  $B_\nu$ \fi}

\def\msigma{\ifmmode M_{\sigma} \else $M_{\sigma}$\fi}
\def\mbulge{\ifmmode M_{\mathrm{bulge}} \else $M_{\mathrm{bulge}}$\fi}
\def\mgal{\ifmmode M_{\mathrm{gal}} \else $M_{\mathrm{gal}}$\fi}
\def\lgal{\ifmmode L_{\mathrm{gal}} \else $L_{\mathrm{gal}}$\fi}
\def\lbulge{\ifmmode L_{\mathrm{bulge}} \else $L_{\mathrm{bulge}}$\fi}
\def\mgalstar{\ifmmode M^*_{\mathrm{gal}} \else $M^*_{\mathrm{gal}}$\fi}

\def\mbhsigstar{\ifmmode M_{\mathrm{BH}} - \sigma_* \else $M_{\mathrm{BH}} - \sigma_*$ \fi}
\def\deltalogmbh{\ifmmode \Delta~{\mathrm{log}}~M_{\mathrm{BH}} \else $\Delta$~log~$M_{\mathrm{BH}}$\fi}

\def\sigstar{\ifmmode \sigma_* \else $\sigma_*$\fi}
\def\sigthree{\ifmmode \sigma_{\mathrm{[O~III]}} \else $\sigma_{\mathrm{[O~III]}}$\fi}
\def\sigtwo{\ifmmode \sigma_{\mathrm{[O~II]}} \else $\sigma_{\mathrm{[O~II]}}$\fi}

\begin{document}

\title{Quasars with a Kick  -- Black Hole Recoil in Quasars\altaffilmark{1}}

\author{G.~A. Shields\altaffilmark{2}, E.~W. Bonning \altaffilmark{3}, 
and S.~Salviander \altaffilmark{2}}

\altaffiltext{1}{To appear in ``Black Holes. Poster Papers from the
  Space Telescope Science Institute Symposium, April 2007," Mario
  Livio and Anton Koekemoer, eds., STScI.}

\altaffiltext{2}{Department of Astronomy, University of Texas, Austin,
TX 78712; shields@astro.as.utexas.edu, triples@astro.as.utexas.edu} 

\altaffiltext{3}{LUTH, Observatoire de Paris, CNRS, Universit\'{e} Paris
 Diderot; Place Jules Janssen 92190 Meudon, France; erin.bonning@obspm.fr}

\begin{abstract}

Mergers of spinning black holes can give recoil
velocities from gravitational radiation up to several thousand \kms.  A
recoiling supermassive black hole in an AGN can retain the inner part of its
accretion disk, providing fuel for continuing AGN activity.   Using AGN in the
Sloan Digital Sky Survey (SDSS) that show
velocity shifts of the broad emission lines relative to the narrow lines, we place upper limits
on the incidence of high velocity recoils in AGN.
Brief but powerful flares in soft X-rays may occur when bound material falls back into
the moving accretion disk.

\end{abstract}

\keywords{galaxies: active --- quasars: general --- black hole physics}

\section{Introduction}
\label{sec:intro}

Recent numerical simulations of binary black hole mergers by a number of
groups show large recoil velocities or ``kicks'' resulting from
from anisotropic emission of gravitational radiation [see, e.g., \cite{campa07b}
for a summary and references].  
These results show a maximum kick for spins anti-aligned and
perpendicular to the orbital angular momentum, reaching $2500~\kmps$
for spin $a_* = 0.8$
\citep{gonzalez07,tichy07}. 
\citet{campa07b} predict a maximum recoil velocity of $4000~\kmps$ for
equal mass black holes with maximal spin.
For  a binary supermassive black hole ($\sim10^8~M_\odot$) formed
during a galactic merger \citep{begelman80}, the kick may displace the black
hole from the galactic nucleus or eject it entirely
\citep[][and  references therein]{merritt04}. 

If an accretion disk is present, the inner part will remain bound to the
black hole, fueling AGN activity as the black hole
leaves the galactic nucleus .    
For a merger taking place in an AGN, the accretion disk
fueling the AGN will remain bound to the recoiling  
black hole inside the radius
 $\rbound = {10^{18.1} M_8}/{v_{1000}^2}$~cm  where the orbital velocity equals
\vkick. Here, $M_8 = \mbh/10^8 M_\odot$ and $v_{1000} =
\vkick/1000$ \kms. For an
$\alpha$-disk \citep{shakura73,frank02}, the disk mass \mbound\ that remains bound is
\begin{equation}
\mbound = (10^{8.0} M_\odot) \alpha^{-4/5}_{-1} M^{3/2}_8
 \dot M^{7/10}_{0} v^{-5/2}_{1000} 
\label{eq:mv}
\end{equation}
where $\dot M_{0}$ is the accretion rate in solar masses per
year \citep{loeb07, bonning07}, subject to $M_{\rm disk} <
M_{\mathrm{BH}}$   for stability. 
Until the disk is depleted, the post-merger accretion rate
will resemble the pre-merger accretion rate, giving a disk consumption time
$t_d ~\approx \mbound/\dot M_0 \approx  (10^8~{\rm yr})
\alpha^{-4/5}_{-1} M^{3/2}_8 \dot M^{-3/10}_{0} v^{-5/2}_{1000}$.
Thus a recoiling AGN could shine for a time comparable with
estimates of the total AGN lifetime.
Here we discuss two observational manifestations
of black hole recoils in AGN, involving Doppler shifts of the emission lines
and energy released by infall into the disk following the recoil.

\section{Shifted emission lines}\label{sec:shift}

The broad 
emission-line region (BLR) of AGN corresponds to material that
remains bound to the black hole, whereas the narrow lines arise  from gas
in the potential of the host galaxy that will not follow the
recoiling black hole \citep{merritt06}. Therefore,
the broad lines will appear shifted with respect
to the galaxy systemic redshift as expressed by the narrow emission lines. 
We have carried out a search for candidate kicked QSOs using spectra
from the Sloan Digital Sky Survey\footnote{The SDSS website is 
http://www.sdss.org.} Data Release 5 (DR5).
We used  objects spectroscopically identified as QSOs by
SDSS having $0.1 < z < 0.81$ such that \hbeta\ 
and \oiii$~\lambda5007$ were both measurable. The lines were measured 
as in \citet{salviander07}, involving subtraction of \feii\ and narrow \hbeta, and 
a Gauss-Hermite fit to the line profiles with asymmetry parameter $h_3$ and
kurtosis parameter $h_4$. Our final 
data set consists of 2598 objects.  This reflects the various quality cuts 
of \citet{salviander07} plus elimination of objects with 
strongly asymmetrical \hbeta\  giving $h_3$
greater than 0.1 in absolute value.  The
displacement  of the broad \hbeta\ line peak with respect to the peak of \oiii\
was calculated as $\dvhb = c(z\,_{\hbeta} - z\,_\mathrm{O~III})/(1+z\,_\mathrm{O~III})$ 
and analogously for \mgii.
For details see \citet{bonning07}.

The distribution of \dvhb\ has a mean displacement of 
+100~\kms\ and a FWHM of  500~\kms.   The distribution of \dvmg\
has a similar FWHM and mean displacement of 
-150~\kms.   
In our data set of 2598 objects, the fractions $f_v$ having  \dvhb\ greater
than a given value are:
$f_{\rm 500} = 0.04$, 
$f_{\rm 800} = 0.007$, 
$f_{\rm 1000} = 0.0035$, 
$f_{\rm 1500} = 0.0012$,
$f_{\rm 2000} = 0.0008$, and
$f_{\rm 2500} = 0.0004$.
However, we regard these values as upper limits on the number of
true recoils: 
(1)  \mgii\ typically has a smaller shift than \hbeta, such that 
$ \dvmg\ \approx 0.6 \dvhb$ .  For recoils, one expects similar
shifts for all broad lines.
(2) The largest shifts occur only for objects with large \hbeta\ FWHM, 
a correlation not expected for recoils.
(3) The high \dvhb\ objects generally have normal
narrow line intensities and 
\nev\ profiles wider than the lower
ionization lines, consistent with a normal NLR with a central ionizing source.
(4)  Some high \dvhb\ objects show a characteristic asymmetry involving
a suppression of the flux on one side of the line, suggesting complications
in the BLR.

Among the 501 objects 
having measured velocities for both \hbeta\ and \mgii, there are no
objects with $\dvmg > 800~\kmps$ for which \dvhb\ and \dvmg\ agree
within expectation.  This suggests $f_{800} < 0.002$, a tighter limit than above.

For mergers of two black holes with spin
$a_* = 0.9$ and a range of mass ratio 
$q \equiv m_2/m_1 > 0.2$, 
\citet{schnitt07} estimate  theoretical values
$f^i_{\rm 500} = 0.31$ and $f^i_{\rm 1000}= 0.079$. 
Convolving with  
random kick inclinations, we find the predicted 
line-of-sight fractions to be $f_{\rm 500} = 0.18$ and $f_{\rm
1000}= 0.054$. 
Our observational results give upper limits on $f_v$ substantially lower than these
predictions.  However, a number of factors may contribute to the low
observed incidence of recoil shifts.  
(1)  A thorough exploration of
parameter space for mergers is needed using full numerical relativity
simulations \citep{campa07b}. 
(2)  The distribution of spin parameters
$a_*$ for supermassive black holes remains uncertain, as does the
expected distribution of merger parameters (mass ratio, spin
orientations). For example, \citet{bogda07} argue that gas accretion
will align the spins with the orbital angular momentum vector, 
resulting in small kicks. 
(3)  The likelihood of
black hole mergers early in the luminous phase of an AGN is uncertain.
(4) The observed shift velocity may be substantially reduced by 
dynamical effects including the inertia of the bound disk and the stars that
remain bound to the recoiling hole, as well as the slowing of the black hole
by the gravitational potential of the galaxy and by dynamical friction over the
luminous life of the wandering AGN \citep{merritt04,madau04}.

These dynamical effects are less severe for large kick velocities.
For our sample, the median value of $\mbh$, derived from the \hbeta\ line
width \citep{salviander07} is
$10^{7.9}~\msun$, with $L/\led \approx 10^{-0.8}$, and $\mdoto = 10^{-0.5}$.  
Here we have assumed
$L = 0.1\mdot c^2$ and $\lbol = 9\nuLnu$ \citep{kaspi00}. 
For this configuration, the bound disk mass is $\mbound \approx
(10^{7.4}~\msun)~\vthree^{-5/2}$.  The momentum of the recoiling black hole will
be shared by the mass of the bound disk traveling with it, giving
$v = \vkick \mbh/(\mbh + \mbound)$. 
However, for $\vkick > 1500~\kmps$ this  does not seriously  
slow the hole.   
The typical \mbh\ for 
our sample corresponds to $\sigstar \approx 175~\kmps$ by the
\mbhsigstar\ relationship of
\citet{trem02}.  
A recoil at $\vkick =
2000~\kmps$ gives $\mbound = 0.07 \mbh$, which slows the hole to
1900~\kmps.  The disk consumption time $t_d$ is about
17~million  years.   We have estimated the slowing of the hole in
a simple model for the host galaxy resembling the isothermal
sphere model of \citet{madau04}.  
The hole reaches a radius of roughly 30~kpc while slowing to $\sim 1500~\kmps$
during the luminous phase.
(For this high velocity,  dynamical friction is minor during $t_d$.)  
Most of the luminous phase is spent
close to this velocity,  and half of the objects will have a
line-of-sight velocity component  above 800~\kmps\ for random orientations.  

If the AGN in our sample have
a typical lifetime of 50 million years, whereas the life of objects with
$\vkick > 2000~\kmps$ is about 17 million years, then the fraction of AGN
with an observed shift over 800~\kms\ will be
$f_{800} \approx (0.5) (17/50) f^i_{2000}$, 
where $f^i_{2000}$ is the intrinsic fraction of recoils over 2000~\kms.  This may
be compared with our observed limit of less than 0.2\% for shifts over
800~\kmps, implying that $f^i_{2000} <  1.2\%$. This example
suggests that the statistics of velocity shifts have potential to give
useful constraints on high velocity recoil events.   

\section{Recoil flares}
\label{sec:flare}

Marginally bound material will lag behind the recoiling hole and
then  fall back into the disk with velocity
$\sim \vkick$.  In the frame of the moving hole, the bound disk material has
excess energy $\eflare = (1/2)\mbound \vkick^2$ compared to material in circular orbit
with the pre-kick angular momentum. 
The energy associated with the mass
given by Eq. \ref{eq:mv} is
\begin{equation}
\eflare =  (10^{57.3}~\erg) \alpha^{-4/5}_{-1}  M^{3/2}_8
  \dot M^{7/10}_{0}  \vthree^{-1/2}.
\label{eq:eflare}
\end{equation}
If this energy is dissipated  on the dynamical timescale  
\begin{equation}
\tflare = \rbound/\vkick =  (420~\yr) M_8 \vthree^{-3}
\label{tflare}
\end{equation}
by shocks as the material falls back, the
power associated with this ``recoil flare''  is 
\begin{equation}
\lflare = (10^{47.2}~ \ergps) \alpha_{-1}^{-4/5} M_8^{1/2} \dot M_0^{7/10} \vthree^{5/2}.
\label{eq:lflare}
\end{equation}
The post-shock temperature will be
\begin{equation}
T_{shock} = (10^{6.9}~\mathrm{K} )(v_{1000})^2 = (0.7~\kev) \vthree^2
\label{eq:tshock}
\end{equation}
 (Osterbrock \& Ferland 2006). 
The flare may substantially exceed
the luminosity of the QSO in its active phase as well as the Eddington limit,
raising the possibility of radiation driven outflows.

The shocks will emit soft X-rays,  but absorption by the infalling material will
be severe.
Consider a QSO with $\mbh = 10^8~\msun$ accreting at $10^{-0.5}~\msunyr$  
(giving $L/\led \approx 10^{-0.9}$) before the final black hole in-spiral.
For a kick velocity of $1000~\kmps$, the bound disk has
mass $10^{7.7}~\msun$, radius
$\rbound = 10^{18.1}~\cm$, and surface density $\Sigma \approx
10^{4.2}~\mathrm{ g~cm^{-2}}$.  This corresponds to an
electron scattering optical depth (if ionized) $\tau_e \approx 10^{3.8}$. 
The optical depth due to photoelectric absorption by H, He, and heavy
elements will be large.  The density of the infalling material is 
$N = 10^{9.9}$ atoms per cubic centimeter.  For a rough estimate of the spectrum
of the cooling shocked gas, 
we computed a coronal equilibrium model
using  version 07.02.00 of the photoionization code CLOUDY, 
most recently described by \citet{ferland98}.   We used $T = 10^7~\kelvin$,
$N = 10^{10.6}~\cmmithree$, and solar abundances.
The cooling time  is $\sim10^{-3.1}~\yr$, with much of the total
cooling in the form of thermal bremsstrahlung.  There are many 
strong emission lines, including lines of
Fe~{\sc XVII} to Fe~{\sc  XX} at 12 to 15~\AA, each with several
percent of the total energy.
The thickness of the cooling zone will be $\sim 10^{11.8}~\cm$,  
small compared with $R_b$.

The X-ray luminosity will photoionize the infalling material approaching the shock.  The local
energy flux in the X-ray flare is $\fflare \approx  10^{9.8}~\ergpspcm$, 
giving an ionizing photon flux of $\phi_i \approx 10^{19.4}~\cmmitwo$.  
We have computed photoionization models with CLOUDY for this
precursor zone.  We used a thermal bremsstrahlung
ionizing spectrum at $10^7~\kelvin$ with flux  $10^{10}~\ergpspcm$,
a gas density $N = 10^{10}~\cmmithree$, and a cutoff
column density $10^{24}~\cmmitwo$.
The photoionized depth 
in this pre-shock flow is $\sim10^{13.5}~\cm$, again narrow
compared to $R_b$.  
The high ionization parameter ($U = 10^{-0.9}$) and hard ionizing spectrum
give strong ultraviolet emission lines of highly ionized species.
The \lalpha\ line has an emergent flux of  8\%\ of the incident flux,
28 times the flux of \hbeta.  Other line intensities include
$I(\lambda)/I(\mathrm{Ly}\alpha) =  1.5$
for O~{\sc VI}~$\lambda1035$ and 0.8 for
C~{\sc IV}~$\lambda1549$.  The high density suppresses forbidden line emission.
The line widths will be $\sim \vkick$.  
Any surrounding dusty torus (Antonucci \& Miller 1985)
will in turn reprocess  much of this
energy into infrared radiation from dust.

The number of recoil flares currently observable
involves the rate of mergers, the probability of kicks of various velocities,
and the flare duration at a given velocity.   Let $dN/dt$ be the merger rate per
year of observer time in some range of \mbh\ and redshift $z$.  Roughly, the number of kicks currently
observable is $\nflare \approx (dN/dt) \, [\tflare (1+z)] \, f^i_v$, where $f^i_v$ is as above.
Haehnelt (1994) derived the black hole merger rate from the
QSO luminosity function.
For $\mbh$ between $10^7$ and  $10^8~ \msun$,  Haehnelt finds 
$0.016$ events per year of observer time
for $z < 3$, giving roughly 6, 0.2 flares in play 
for $\vkick > 500, 1000~\kmps$, respectively.    
Uncertainties include  the flare duration,
the number of obscured QSOs, and Haehnelt's  assumed $L/\led$ and QSO lifetime.

For a $10^8~\msun$ hole in a QSO shining at $0.3\led$ at $z = 2$, the received flux
from the flare is $\sim 10^{-11.7}~\ergpspcm$ for $\vthree = 1$. 
The observed spectrum will be softened by the
redshift, but a considerable fraction of the radiation should still be received at photon
energies above 0.2 keV.   At $10^{-11.7}~\ergpspcm$, 
Hasinger, Miyaji, \& Schmidt (2005) find 
$\sim 0.2$ soft X-ray sources per square degree, 
so that the recoil flares would need to be
identified from among roughly $10^4$ other sources of similar flux.

Recoil flares may provide a unique opportunity to detect
high velocity recoils in AGN.  Detailed simulations of the gas dynamics
and emitted spectrum are needed to help identify these rare events.

\acknowledgments
We thank  Ski Antonucci, Omer Blaes, 
Richard Matzner, Meg Urry, and Marta Volonteri for helpful discussions and communications. 
EWB is supported by Marie Curie Incoming European Fellowship contract
MIF1-CT-2005-008762 within the 6th European Community Framework
Programme.

\end{document}
